# Inducing superconductivity in Weyl semi-metal microstructures by selective ion sputtering

Short title: Irradiated Weyl nanowires go superconducting

One sentence summary: A Niobium sheet forms on NbAs crystals under ion bombardment, leading to a simple fabrication route towards Weyl semi-metal microstructures with proximity-induced superconductivity.

**Authors**

Maja D. Bachmann[1,2], Nityan Nair[3], Felix Flicker[3], Roni Ilan[3,4], Tobias Meng[5], Nirmal J. Ghimire[6], Eric D. Bauer[6], Filip Ronning[6], James G. Analytis[3], Philip J.W. Moll[1*]

**Affiliations**

[1]Max-Planck-Institute for Chemical Physics of Solids, 01187 Dresden, Germany.

[2]Scottish Universities Physics Alliance, School of Physics and Astronomy, University of St. Andrews, St. Andrews KY16 9SS, United Kingdom.

[3]Department of Physics, University of California Berkeley, CA, USA.

[4]Raymond and Beverly Sackler School of Physics and Astronomy, Tel Aviv University, Tel Aviv 69978, Israel.

[5]Institut für Theoretische Physik, Technische Universität Dresden, 01062 Dresden, Germany.

[6]Los Alamos National Laboratory, Los Alamos, NM, USA.

*correspondence address: philip.moll@cpfs.mpg.de

**Abstract**

By introducing a superconducting gap in Weyl- or Dirac semi-metals, the superconducting state inherits the non-trivial topology of their electronic structure. As a result, Weyl superconductors are expected to host exotic phenomena such as non-zero-momentum pairing due to their chiral node structure, or zero-energy Majorana modes at the surface. These are of fundamental interest to improve our understanding of correlated topological systems, and moreover practical applications in phase coherent devices and quantum applications have been proposed. Proximity-induced superconductivity promises to allow such experiments on non-superconducting Weyl semi-metals. Here we show a new route to reliably fabricating superconducting microstructures from the non-superconducting Weyl semi-metal NbAs under ion irradiation. The significant difference in the surface binding energy of Nb and As leads to a natural enrichment of Nb at the surface during ion milling, forming a superconducting surface layer ($T_c$~3.5K). Being formed from the target crystal itself, the ideal contact between the superconductor and the bulk may enable an effective gapping of the Weyl nodes in the bulk due to the proximity effect. Simple ion irradiation may thus serve as a powerful tool to fabricating topological quantum devices from mono-arsenides, even on an industrial scale.

**Introduction**

Materials with non-trivial band structure topologies form one of the most active fields of current condensed matter research[1-3]. Their unifying feature is the existence of topologically protected surface states: two-dimensional Dirac particles on the surface of topological insulators, or open strings of surface states known as "Fermi arcs" in the case of three-dimensional topological semi-metals. When these topological surfaces are superconducting, they may host zero-energy modes related to "Majorana fermions", proposed as one route to topological quantum computation[4-6]. The key ingredient is a non-zero superconducting order parameter breaking charge conservation on the surface, which can be achieved either by an intrinsic superconducting gap in the bulk, or by inducing superconductivity through coupling the surface to a superconductor. The latter opens exciting possibilities to engineer Majorana systems by combining band-inverted, strong spin-orbit coupled topological materials with robust superconductivity, without the need to obtain both intrinsically within the same material.

In the past few years, substantial effort has been devoted to the engineering of proximity-induced topological superconductivity in the surface states of topological insulators[7–22]. Since these materials are gapped in the bulk by definition, the proximity effect can only be active on their surface. A fundamentally different situation is found in topological semi-metals. In this paper, we focus on topological Weyl semi-metals (WSMs) featuring an energy gap for all momenta except a small number of topologically-protected "Weyl nodes"(isolated points in the Brillouin zone at which the gap closes). Weyl semi-metals require a breaking of either time-reversal or inversion symmetry. While theoretical studies have focused mostly on the former, it is examples of the latter which have been experimentally confirmed, in the mono-arsenide materials (Ta,Nb)(As,P). NbAs is one of the best-characterized[23], with its topological character having been identified by Angle Resolved Photoemission Spectroscopy (ARPES)[24], transport[25,26], and high-field magnetization[27]. Superconductivity in WSMs, both intrinsic and extrinsic, has been studied theoretically in recent years, but has not yet been realized experimentally.

The WSM NbAs does not exhibit intrinsic superconductivity, even under high hydrostatic pressure[41]. Methods aimed at changing the charge carrier concentration in the hope of inducing superconductivity, such as gating or chemical doping, have so far not been applied successfully without destroying the topological character of the material[42]. A natural route to overcome this issue is to induce superconductivity via the proximity effect in NbAs. Such an experimental configuration is typically achieved by deposition of a superconducting thin film onto the crystalline sample (either bulk or thin film). This approach has enabled the study of proximity-effect-induced superconductivity on topological insulators in two and three dimensions, as well as one-dimensional systems with strong spin orbit coupling in geometries such as Josephson junctions and normal-superconducting interfaces[23,43–46].

One outstanding obstruction on the path to a well-developed proximity-induced gap arises from the interface quality that commonly suffers from lattice mismatches and the unwanted formation of interface layers when switching between different processing materials or exposing the material to the atmosphere. Here we present an alternative, technologically simple and effective way to fabricating superconductor-WSM heterostructures reliably. TaAs and NbAs contain elemental superconductors with sizable transition temperatures Ta ($T_c$=4.5K) and Nb ($T_c$=9.2K). Besides the intrinsic superconductivity of these elements, their binary and tertiary compounds show a rich phase space of superconducting states with high transition temperatures, with $Nb_3Sn$ and NbTi being the most famous and commercially most important examples. This material-specific affinity to superconductivity provides the opportunity of an alternate approach to induce surface superconductivity by removing As from the surface instead of adding a superconductor onto it. If the surface of a crystal can be locally depleted of As, an intrinsically superconducting thin film would form as a shell around the crystal.

**Results**

Low energy ion irradiation can be used to effectively remove As from an approximately 20nm deep surface layer in single crystals of NbAs, and induce robust type-II superconductivity with a $T_c \sim 3K$. By using a Focused Ion Beam (FIB), selected regions of a NbAs sample can be locally irradiated, providing a simple method to form superconducting nanostructures in natural proximity to the Weyl semi-metal host material. This process could therefore be used to selectively turn parts of a NbAs thin film circuit superconducting. Typical acceleration voltages of 30kV for $Ga^{2+}$ are used in the FIB process[47], which is found to be highly effective for As depletion. During ion irradiation, the impacting ions transfer their energy to the target and there is a probability for the target atoms to escape the solid, i.e. to be sputtered. The energetics of the escape process is dominated by the surface binding energy of the constituent atoms, which can be shown to be well correlated with the heat of sublimation. Fortuitously, NbAs is a binary compound consisting of two elements with extreme sublimation points: while As sublimates at a relatively low temperature of 887K (34.76 kJ/mol), Nb is well-known for its high boiling point of 5017K (689.9 kJ/mol). This energetic difference already captures the essential physics of the natural self-enrichment of Nb on the surface of NbAs under ion bombardment. To further quantify this intuition, the sputter yield ratios were calculated using the SRIM-2008[48] Monte-Carlo based code simulating the Ga ion impact interactions at the complete damage cascade level. At normal incidence for 60keV Ga ions, the sputter yield is 1.70 per incident ion for Nb, and 10.01 for As. Therefore one expects a highly Nb-rich amorphous surface with a composition of $Nb_{5.88}As$ to form under sputtering equilibrium conditions, as suggested by the difference in the heat of sublimation.

Figure 1A shows a typical NbAs microstructure fabricated by FIB micromachining to study the irradiation-induced surface superconductivity. The final device is a crystal micro-bar with electric contacts in a four-terminal resistance configuration. During fabrication, the bar is equally irradiated from all sides to ensure that it is completely encased in a shell of ion-beam irradiated material, as shown in the sketch in figure 1B. The detailed fabrication recipe is given in the methods section. Typical cool down curves for crystal bars of different cross-sections are shown in figure 2A. The weak metallic temperature dependence is typical for semi-metals. The absolute resistance of the devices cannot be explained using a simple geometric factor and a constant specific resistivity. This failure of Ohmic scaling directly evidences the presence of the conductive surface layer and its role in the charge transport. More than 10 microstructures of different dimensions were fabricated, and zero-resistance superconductivity with a well reproducible $T_c \sim 3K$ was observed in all of them. The method of ion surface treatment is equally effective in all members of the (Ta,Nb)(As,P) mono-arsenide family, given the large chemical and physical similarity between (Ta,Nb) and (As,P). Accordingly, they all display superconducting transitions with $T_c \sim 2-4K$ (Figure 2B).

**Discussion**

A key aspect of proximity-effect-induced superconductivity concerns the quality of the interface between the superconductor and the metal. Traditionally, such heterostructures are fabricated by deposition of a superconducting metal on a non-trivial material and thus often suffer from non-ideal interfaces due to surface contamination. Therefore such structures are commonly described by a diffusive coupling of the superconductor into the metal in the dirty limit, where the range of superconductivity in the metal is limited by the short mean-free-path between defects as described by the Usadel equations[49]. The process described here leads to a self-formation of the SC-normal metal interface well within the crystalline target material, which potentially enhances the transparency of the interface and thus extends the range of the proximity coupling – key to the fabrication of microstructures with a fully developed gap in the bulk.

The resistance of the microwires is temperature independent below 10K, and the large upper critical fields suggest a robust type-II superconducting state (figure 3). The field dependence in the normal state (figure 3C) provides further insight into the microscopic details of the current flow in the structures. At low fields, a quasi-quadratic magnetoresistance is observed, followed by a successive saturation to a slow increase. This behavior can be well understood by the simple assumption of two conductive pathways in the sample, one through the pristine NbAs bulk, and one through the Nb-rich layer on the surface as depicted in figure 1. Assuming a usual magnetoconductivity $\sigma_{NbAs}=(a\ H^2+const.)^{-1}$ for the NbAs metal and a field-independent conductivity for the disordered Nb-layer, $\sigma_{Nb}$, we arrive at a simple expression for the magnetoconductivity $\sigma=\sigma_{NbAs}+\sigma_{Nb}$, which well-describes the experimental field dependence (Fig.3c). The fit to the data yields a resistance of the surface Nb-film of 27.8Ω for this 300nm wide, 2.7μm tall, and 10μm long FIB structure. This resistance of the Nb layer agrees quantitatively with measurements of FIB-irradiated Nb-thin films[50]. The fact that a simple model assuming only two conduction channels explains the data well is further evidence for the presence of a well-defined Nb-layer on the surface, instead of a gradual increase of the Nb concentration over an extended surface region.

The extent the superconducting region reaches into the topological semi-metal via the proximity effect is an important aspect to fabricating topological microdevices. Figure 4 shows the current-voltage characteristics of a NbAs microwire at different widths. These data are taken on the same microwire, which was successively thinned down by FIB milling after each I-V measurement, to exclude any device-dependent issues. A well-defined critical current is observed at each width step, and a clear trend of the critical current is shown in the inset and figure 4B. The critical current depends approximately linearly on the device width. For a homogeneous bulk superconductor thinner than the magnetic penetration length, the critical current is expected to scale with the device width, i.e. $I_c\ (w)=j_c\ wh$ where h denotes the constant height of the sample during thinning, and w is the lateral width varied by the thinning experiments. Trivially, a zero critical current arises at zero width in this case, which appears to be at odds with the present data suggesting a finite intercept. A finite intercept, however, would be natural for superconductivity in a thin surface layer around a metal. Ignoring a potential contribution of proximity-effect induced superconductivity in the bulk, an infinitesimally thin superconducting surface layer may be assumed, leading to a trivial scaling of the critical current in the heterostructure: As all currents flow exclusively on the surface, the critical current density should be proportional to the circumference, $I_c\ (w)=\alpha\ (2h+2w)$ where α is a parameter describing the critical current per unit length of the surface film. At zero width of the metallic core a finite critical current is expected, $I_c\ (0)=2\alpha h$, due to the currents carried by the infinitesimally thin sidewalls.

The width scaling thus supports a superconducting shell scenario, yet at the same time is consistent with supercurrents in an extended proximity-effect induced layer of finite thickness in the metallic bulk. The fabrication conditions for top- and side-walls are identical in terms of ion current density, acceleration voltage and grazing incidence angle. Thus it is natural to assume that the critical current parameter α does not change between top- and side-surface. Using the experimentally estimated intercept, $I_c(0)=164\mu A$, and the measured device height $h=2.7\mu m$, the only free parameter in this scenario can be estimated, $\alpha=30.4\mu A/\mu m$. Such a model of supercurrents confined to the thin film on the surface, $I_c(w)=60.7\mu A/\mu m\ (h+w)$ matches the experimental intercept by design, but it underestimates the experimentally observed slope (blue dashed line). This suggests that the critical current grows faster than the circumference of the device, which would be a natural consequence of supercurrents flowing in an extended region in the microwire bulk. The next step towards confirming topological superconductivity in these structures will be to integrate the superconducting microwires into phase-coherent heterostructures, such as Josephson junctions. Such experiments probe the superconducting phase gradient in a topological superconductor directly, and would provide the strongest evidence for superconductivity in the NbAs core via the appearance of a half-flux quantum in the Fraunhofer pattern.

The results presented here establish robust superconductivity in FIB-prepared microstructures of NbAs. At the same time, the irradiation with Ga ions is known to be associated with a deterioration of the material quality and could potentially have destroyed the materials crystal structure. The resulting amorphous bar could hence be a robust but topologically trivial superconductor. However, the experiment self-consistently evidences that the FIB-prepared microstructures of NbAs retain the high quality of the parent crystal beneath the 20nm deep Nb-rich shell. Figure 5 shows the magnetoresistance at 2K in the normal state, showing pronounced Shubnikov-de Haas oscillations. These oscillations in the density of states arise from quantum coherent motion of quasiparticles around the Fermi surface, and thus are well known to be exponentially suppressed even at small defect densities[51]. Their observation is a clear indication that the crystalline symmetry is well preserved in the devices. At the same time, it suggests that a well-defined Fermi-surface exists in the core, evidencing that the region of As deficiency is indeed confined to a thin defect layer encasing the structure. The observed frequency of 85(+/-1.5) T is in excellent agreement with previous reports of de Haas-van Alphen oscillations at 84T obtained from bulk crystals for the given field configuration (HIIa)[26,27]. This quantitative agreement evidences the unchanged position of the chemical potential compared to bulk crystals, thereby excluding a Lifshitz transition to a topologically trivial state due to effective charge carrier doping during the irradiation process.

The fabrication of proximity-effect-induced superconducting heterostructures is exceedingly demanding on the purity of the starting materials and the cleanliness of the fabrication process, as even slight contaminations of the interface between the materials forms an effective barrier strongly suppressing the Cooper pair tunneling amplitude – a well-known complication for the industrial fabrication of superconducting junctions such as Superconducting Quantum Interference Devices (SQUIDs). Our process of ion-irradiation of a NbAs/TaAs thin films requires only a single material synthesis step and the interface between the topological metal and the superconductor is formed within the material itself. Low voltage, broad band ion irradiation such as Ar-ion sputtering is a standard technique in chip fabrication, and can easily be combined with photo- or electron-lithographic techniques to selectively turn parts of a circuit superconducting. We thus envision this process to be a promising route to fabricating topological superconductor heterostructures on a large scale.

## Materials and Methods

### Ion beam surface treatment

All studied samples were fabricated in a FEI Helios Nanolab Focused Ion beam, and an acceleration voltage of 30kV was used. To ensure a homogeneous coating of the microwires with the Nb-enriched surface layer, a sample fabrication scheme was developed that exposed all surfaces equally to the ion irradiation. The procedure starts on a large, flat growth face of a single crystal of NbAs.

In a first step, a 120 x 20 x 3 $\mu m^3$ slice was cut from a mm-sized single crystal of NbAs and transferred to a Si/SiO$_2$ substrate and mounted in an epoxy droplet. Electrical contacts were fabricated by direct gold evaporation followed by an ion milling step to remove the gold from the main device. In the second step, the device was reintroduced into the FIB chamber and fine-structured into the final form shown in figure 1 using an intermediate current of 800pA. Finally, the sidewalls of the structure were fine-polished to the target width using a fine current of 80pA.

## Crystal growth

The growth of NbAs crystals used for this study has been reported previously[25]. Single crystals of NbAs were grown by vapor transport using iodine as the transport agent. First, polycrystalline NbAs was prepared by heating stoichiometric amounts of Nb and As in an evacuated silica ampoule at 700 °C for 3 days. Subsequently, the powder was loaded in a horizontal tube furnace in which the temperature of the hot zone was kept at 950 °C and that of the cold zone was ≈850 °C. Several NbAs crystals formed with distinct well faceted flat plate like morphology. The crystals of NbAs were verified by checking (00l) reflections on a x-ray diffractometer and by compositional analysis conducted using energy dispersive x-ray spectroscopy (EDS). An atomic percentage ratio of Nb:As = 53:47 was obtained on the EDS measurements, which is close to the expected uncertainty of 3–5%.

## References and Notes

1. Wan, X., Turner, A. M., Vishwanath, A. & Savrasov, S. Y. Topological semi-metal and Fermi-arc surface states in the electronic structure of pyrochlore iridates. Phys. Rev. B 83, 205101 (2011).

2. Qi , X.-L. & Zhang, S.-C. Topological insulators and superconductors. Rev. Mod. Phys. 83, 1057 (2011).

3. Bernevig, B. A. It's been a Weyl coming. Nat. Phys. 11, 698–699 (2015).

4. Xu, J. P. et al. Artificial topological superconductor by the proximity effect. Phys. Rev. Lett. 112, 1–5 (2014).

5. Beenakker, C. W. J. W. J. Search for Majorana fermions in superconductors. Annu. Rev. Condens. Matter Phys. 4, 15 (2013).

6. Alicea, J. New directions in the pursuit of Majorana fermions in solid state systems. Rep. Prog. Phys. 75, 76501 (2012).

7. Fu, L. & Kane, C. Superconducting Proximity Effect and Majorana Fermions at the Surface of a Topological Insulator. Phys. Rev. Lett. 100, 96407 (2008).

8. Zhang, D. et al. Superconducting proximity effect and possible evidence for Pearl vortices in a candidate topological insulator. Phys. Rev. B 84, 165120 (2011).

9. Veldhorst, M. et al. Josephson supercurrent through a topological insulator surface state. Nat. Mater. 11, 417–421 (2012).

10. Yang, F. et al. Proximity-effect-induced superconducting phase in the topological insulator Bi 2Se 3. Phys. Rev. B 86, 134504 (2012).

11. Zareapour, P. et al. Proximity-induced high-temperature superconductivity in the topological insulators Bi2Se3 and Bi2Te3. Nat. Commun. 3, 1056 (2012).

12. Williams, J. R. et al. Unconventional Josephson effect in hybrid superconductor-topological insulator devices. Phys. Rev. Lett. 109, 56803 (2012).

13. Kriener, M., Segawa, K., Ren, Z., Sasaki, S. & Ando, Y. Bulk superconducting phase with a full energy gap in the doped topological insulator CuxBi2Se3. Phys. Rev. Lett. 106, 127004 (2011).

14. Sacépé, B., Oostinga, J. B., Li, J., Ubaldini, A. & Nuno, J. G. Gate-tuned normal and superconducting transport at the surface of a topological insulator. Nat. Commun. 2, 575 (2011)

15. Koren, G., Kirzhner, T., Lahoud, E., Chashka, K. B. & Kanigel, A. Proximity-induced superconductivity in topological Bi 2Te 2Se and Bi 2Se 3 films: Robust zero-energy bound state possibly due to Majorana fermions. Phys. Rev. B 84, 224521 (2011).

16. Oostinga, J. B. et al. Josephson supercurrent through the topological surface states of strained bulk HgTe. Phys. Rev. X 3, 21007 (2013).

17. Sochnikov, I. et al. Nonsinusoidal current-phase relationship in josephson junctions from the 3D topological insulator HgTe. Phys. Rev. Lett. 114, 66801 (2015).

18. Kurter, C., Finck, a. D. K., Hor, Y. S. & Van Harlingen, D. J. Evidence for an anomalous current-phase relation in topological insulator Josephson junctions. Nat. Commun. 6, 7130 (2013).

19. Xu, J. P. et al. Experimental detection of a Majorana mode in the core of a magnetic vortex inside a topological insulator-superconductor Bi2Te3/NbSe2 heterostructure. Phys. Rev. Lett. 114, 17001 (2015).

20. Sochnikov, I. et al. Direct measurement of current-phase relations in superconductor/topological insulator/superconductor junctions. Nano Lett. 13, 3086–3092 (2013).

21. Finck, A. D. K., Kurter, C., Hor, Y. S. & Van Harlingen, D. J. Phase coherence and andreev reflection in topological insulator devices. Phys. Rev. X 4, 41022 (2014).

22. Hart, S. et al. Induced superconductivity in the quantum spin Hall edge. Nat. Phys. 10, 638–643 (2014).

23. Nadj-Perge, S. et al. Observation of Majorana fermions in ferromagnetic atomic chains on a superconductor. Science 346, 602–607 (2014).

24. Huang, S.-M. et al. A Weyl Fermion semi-metal with surface Fermi arcs in the transition metal monopnictide TaAs class. Nat. Commun. 6, 7373 (2015).

25. Ghimire, N. J. et al. Magnetotransport of single crystalline NbAs. J. Phys. Condens. Matter 27, 152201 (2015).

26. Luo, Y. et al. Electron-hole compensation effect between topologically trivial electrons and nontrivial holes in NbAs. Phys. Rev. B 92, 205134 (2015).

27. Moll, P. J. W. et al. Magnetic torque anomaly in the quantum limit of Weyl semi-metals. Nat. Commun. 7, 12492 (2016).

28. Schnyder, A. P., Brydon, P. M. R. & Timm, C. Types of topological surface states in nodal noncentrosymmetric superconductors. Phys. Rev. B 85, 24522 (2012).

29. Lu, B., Yada, K., Sato, M. & Tanaka, Y. Crossed surface flat bands of weyl semi-metal superconductors. Phys. Rev. Lett. 114, 96804 (2015).

30. Yang, S. A., Pan, H. & Zhang, F. Dirac and Weyl superconductors in three dimensions. Phys. Rev. Lett. 113, 46401 (2014).

31. Zhou, T., Gao, Y. & Wang, Z. D. Superconductivity in doped inversion-symmetric Weyl semi-metals. Phys. Rev. B 93, 94517 (2016).

32. Bednik, G., Zyuzin, A. A. & Burkov, A. A. Superconductivity in Weyl metals. Phys. Rev. B 92, 35153 (2015).

33. Cho, G. Y., Bardarson, J. H., Lu, Y. M. & Moore, J. E. Superconductivity of doped Weyl semi-metals: Finite-momentum pairing and electronic analog of the 3He-A phase. Phys. Rev. B 86, 214514 (2012).

34. Hosur, P., Dai, X., Fang, Z. & Qi, X. L. Time-reversal-invariant topological superconductivity in doped Weyl semi-metals. Phys. Rev. B 90, 1–9 (2014).

35. Wei, H., Chao, S. P. & Aji, V. Long-range interaction induced phases in Weyl semi-metals. Phys. Rev. B 89, 235109 (2014).

36. Wei, H., Chao, S.-P. & Aji, V. Odd-parity superconductivity in Weyl semi-metals. Phys. Rev. B 89, 14506 (2014).

37. Wang, F. & Lee, D. H. Topological relation between bulk gap nodes and surface bound states: Application to iron-based superconductors. Phys. Rev. B 86, 94512 (2012).

38. Meng, T. & Balents, L. Weyl superconductors. Phys. Rev. B 86, 54504 (2012).

39. Khanna, U., Kundu, A., Pradhan, S. & Rao, S. Proximity-induced superconductivity in Weyl semi-metals. Phys. Rev. B 90, 195430 (2014).

40. Chen, A. & Franz, M. Superconducting proximity effect and Majorana flat bands in the surface of a Weyl semi-metal. Phys. Rev. B 93, 201105(R) (2016).

41. Luo, Y., Ghimire, N. J., Bauer, E. D., Thompson, J. D. & Ronning, F. 'Hard' crystalline lattice in the Weyl semi-metal NbAs. J. Phys. Condens. Matter 28, 55502 (2016).

42. Arnold, F. et al. Negative magnetoresistance without well-defined chirality in the Weyl semi-metal TaP. Nat. Commun. 7, 11615 (2016).

43. Mourik, V. et al. Signatures of Majorana Fermions in Hybrid Superconductor-Semiconductor Nanowire Library. Science 336, 1003–1007 (2012).

44. Das, A. et al. Zero-bias peaks and splitting in an Al-InAs nanowire topological superconductor as a signature of Majorana fermions. Nat. Phys. 8, 887–895 (2012).

45. Deng, M. T., Yu, C. L., Huang, G. Y., Larsson, M. & Xu, H. Q. Anomalous Zero-Bias Conductance Peak in a Nb − InSb Nanowire − Nb Hybrid Device. Nano Lett. 12, 6414–6419 (2012).

46. Chang, W. et al. Hard gap in epitaxial semiconductor-superconductor nanowires. Nat. Nanotechnol. 10, 232–6 (2015).

47. Gianuzzi, L. A. & Stevie, F. A. Introduction to Focused Ion Beams. (Springer Science+Business, 2005).

48. Ziegler, J. F. SRIM-2003. Nucl. Instr. Meth. Phys. Res. B 220, 1027–1036 (2004).

49. Usadel, K. D. Generalized diffusion equation for superconducting alloys. Phys. Rev. Lett. 25, 507–509 (1970).

50. De Leo, N. et al. Thickness Modulated Niobium Nanoconstrictions by Focused Ion Beam and Anodization. IEEE Trans. Appl. Supercond. 26, (2016).

51. Shoenberg, D. Magnetic oscillations in metals. (Cambridge University Press, 1984).

**Acknowledgments**

**General:** We thank Andrew Potter, Andy Mackenzie and Michael Baenitz for stimulating discussions.

**Funding:** MDB and PJWM acknowledge funding through the Max-Planck-Society. MDB acknowledges studentship funding from EPSRC under grant nos. EP/I007002/1. NN is supported by the NSF Graduate Research Fellowship Program under Grant No. DGE 1106400. Work by NN and JGA is partly supported by the Office of Naval Research under the Electrical Sensors and Network Research Division, Award No. N00014-15-1-2674. Work by NN and JGA is partly supported by the Gordon and Betty Moore Foundation's EPiQS Initiative through Grant GBMF4374. FF acknowledges support from a Lindemann Trust Fellowship of the English Speaking Union. RI is funded by AFOSR MURI. TM is funded by Deutsche Forschungsgemeinschaft through GRK 1621 and SFB 1143. NJG and EDB were supported under the auspices of the U.S. Department of Energy, Office of Science. FR was supported by the Los Alamos National Laboratory LDRD program.

**Author contributions:** PJWM and JGA designed the experiment. MDB, NN fabricated the microstructures and performed the transport experiments. FF, RI, TM contributed to the theory and models of proximity effect induced superconductivity in Weyl semi-metals. NJG, EDB, FR, NN, JGA grew and characterized the crystals. All authors were involved in writing the manuscript.

**Competing interests:** The authors declare that they have no competing interests.

**Figures and Tables**

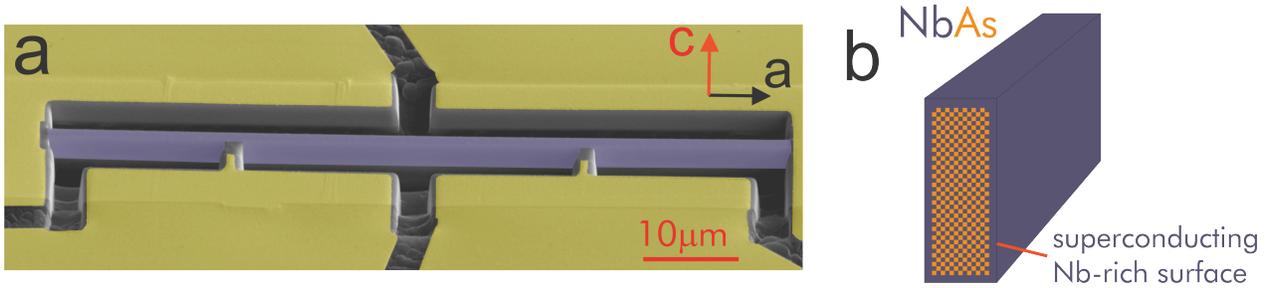

**Fig. 1. Superconducting Weyl Microwires.** (**A**) Scanning electron micrograph of a typical microwire. A $100\times20\times3$ $\mu m^3$ slice was FIB-cut from a NbAs single crystal and electrically contacted (gold leads for 4-terminal measurement). The central bar (purple) was carved in a subsequent FIB etching step, to ensure a homogeneous coverage of the induced superconducting layer. The bar is 2.7 $\mu m$ tall, 1.8 $\mu m$ wide, and the voltage contacts are 35 $\mu m$ apart. (**B**) Sketch of the FIB fabricated wire in cross-section. Arsenic is preferentially sputtered from the surface, leaving a Nb-rich surface layer. The device is completely encased in the superconducting shell.

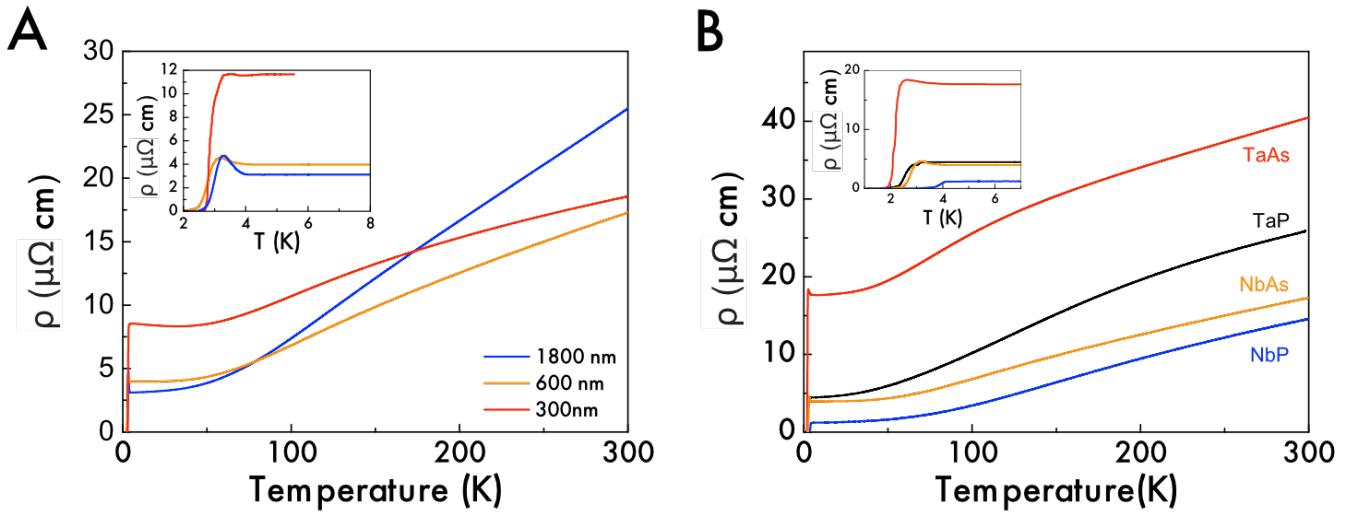

**Fig. 2. Temperature dependence of resistivity.** (**A**) Resistivity as a function of temperature for the NbAs transport bar shown in figure 1. The device was successively thinned down in three steps to widths of 1800nm, 600nm and 300nm. In this process, the height of 2700nm did not change. The device shows a similar onset of the superconducting transition independent of the device width at 3.5K and a zero resistance state below 2.5K. The high temperature resistivity is found to be strongly dependent on the device width, which reflects the presence of a second conduction channel on the surface and the resulting breakdown of Ohmic scaling of the total resistance with the device cross-section. (**B**) Comparison of the temperature dependent resistivity of FIB-microstructured devices across the mono-arsenide family (Ta,Nb)(As,P). Due to the self-formation of a Nb resp. Ta outer shell, all these devices exhibit superconductivity with a $T_c$ between 2 and 4 Kelvin.

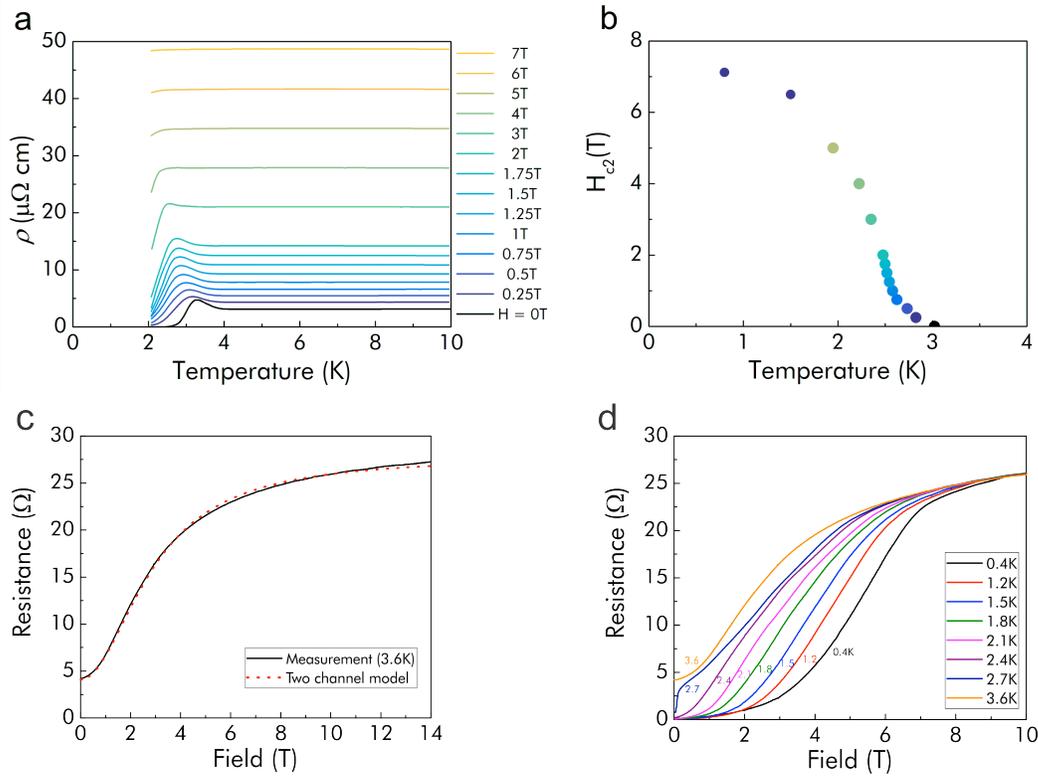

**Fig. 3. Robust superconductivity in magnetic fields.** (**A**) Temperature dependence of the resistivity in magnetic fields applied parallel to the bar (1.8x2.7μm cross-section) and along the crystallographic a-direction. The resistance peak above $T_c$ shrinks and moves to lower temperatures, as superconductivity is suppressed under increasing magnetic fields. However, even in fields of 7T a clear sign of the transition is visible above 2K. (**B**) The upper critical field estimated at the onset of conductivity enhancement through the superconducting transition. To avoid artefacts from the non-monotonic temperature dependence, the upper critical field $H_{c2}(T)$ was defined at 95% of the normal state resistance, i.e. $\rho(T,H_{c2}(T)) = 0.95 \; \rho_n(6K,H_{c2}(T))$. (**C**) Resistance of a 300nm wide NbAs microwire for fields applied perpendicular to the wire in the normal state. The field dependence of the device resistance can be well explained by a two-conductor model, further supporting the presence of a Nb thin film on the device surface. (**D**) Field dependence at various temperatures below $T_c$. Even at temperatures close to $T_c$, large fields are required to completely suppress superconductivity and reach the two-conductor shape of the resistance.

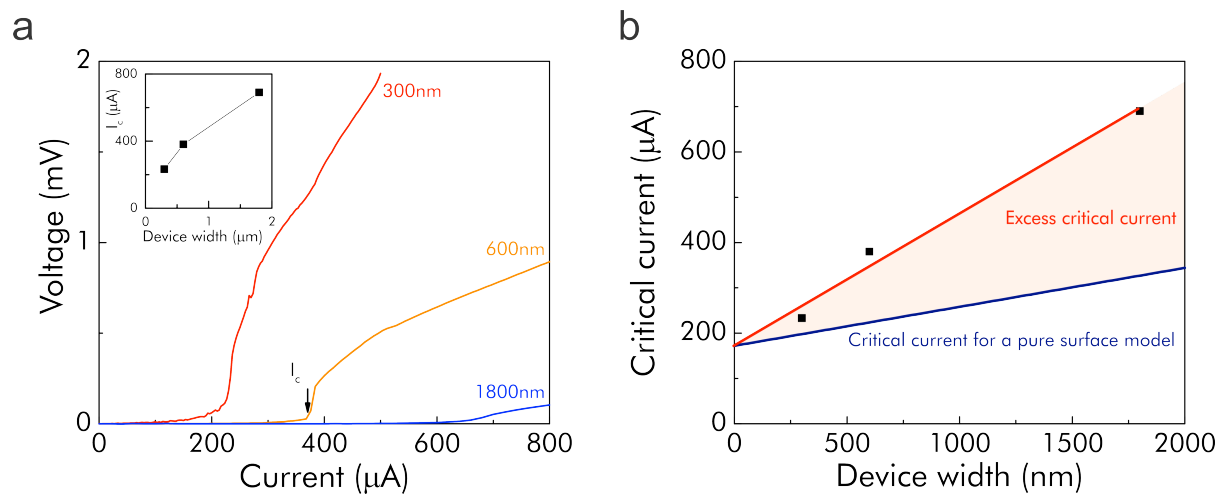

**Fig. 4. Critical currents.** (**A**) Current-voltage characteristic of NbAs crystal bars at different thickness but identical width. For all devices, a sharp increase in voltage signals the destruction of the superconducting state at the critical current $I_c$. Inset: critical current for different device width. (**B**) Comparison of the measured width dependence of the critical currents with a model based on superconductivity confined to the surface. The critical current grows stronger with increasing width than expected from a pure surface model, suggesting that the NbAs bulk plays a role in carrying supercurrents.

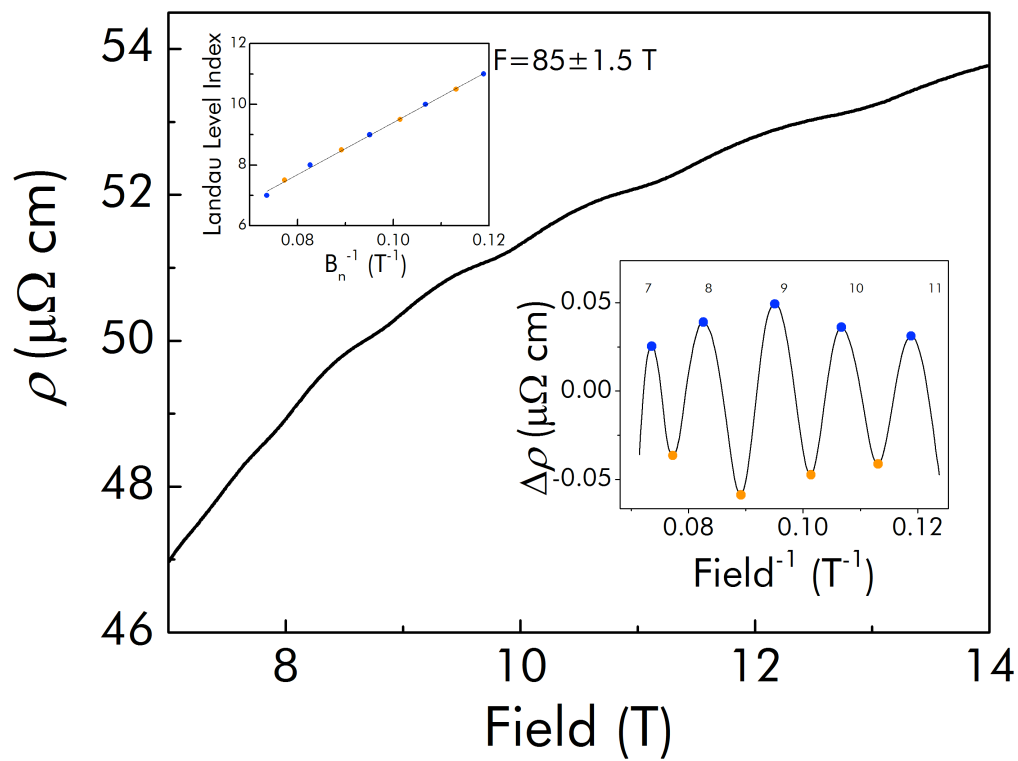

**Fig. 5. Quantum oscillations indicate bulk Fermiology of NbAs.** Magnetoresistance of the 300nm device at 2K in the normal state, for fields along the crystallographic a-direction. Clear quantum oscillations appear on top of the magnetoresistance, even on the thinnest and hence potentially most damaged device. Inset (**right**): same data after background subtraction, highlighting the periodic modulation of the resistivity in inverse field. Inset (**left**): the minima (orange) and maxima (blue) of the resistivity against the Landau level index. The straight line indicates the periodicity in inverse field, and the slope is given by the Fermi surface cross-section of 85T (+/-1.5T).